\documentclass[aps,prd,twocolumn,groupedaddress,showpacs,showkeys]{revtex4-1}
\usepackage{graphicx}
\usepackage{dcolumn}
\usepackage{bm}
\usepackage{amssymb,amsmath,latexsym,amsfonts}
\begin{document}


\title{Elementary Excitations in a BEC with Isotropic Harmonic Trap: Bogoliubov Equations versus Hydrodynamic Formalism}

\author{A. Camacho}
 \email{acq@xanum.uam.mx} \affiliation{Departamento de F\'{\i}sica,
 Universidad Aut\'onoma Metropolitana--Iztapalapa\\
 Apartado Postal 55--534, C.P. 09340, M\'exico, D.F., M\'exico.}


 \date{\today}

 \begin{abstract}
The elementary excitations for a BEC trapped by means of an
isotropic harmonic oscillator are studied in the present work. The
analysis of these perturbations is done in the context of the
Bogoliubov equations and not resorting to the hydrodynamic version.
The comparison between these two approaches will allow us to deduce
a parameter explaining the role that the scattering length and the
trap play in the way in which the frequency of the elementary
excitations acquires information about the angular momentum of the
corresponding solutions. It will be shown that outside the validity
realm of the Thomas--Fermi approximation the frequencies of the
perturbations cannot inherit the information of the angular momentum
codified in the functions describing the elementary excitations.
\end{abstract}

\maketitle

\date{\today}

\section{\bf Introduction}

The analysis of the properties of elementary
excitations in a Bose--Einstein Condensate (BEC) appears as an
important aspect in the study of the properties of a BEC
\cite{Stringari1}. The reasons behind this interest range from the
role played by collective modes and elementary excitations in the
determination of thermodynamic properties \cite{Pethickbook} to the
understanding of the phenomenon of superfluidity \cite{Raman1},
among other possibilities. The analysis of these excitations offers,
at least two options, namely, the hydrodynamic procedure
\cite{Uedabook}, or the formalism known as Bogoliubov equations
\cite{Bogoliubov1}.

At this point we may wonder what kind of logical relation connects
these two procedures, i.e., which one of them is more general, etc.
Bogoliubov case can be used in the analysis of long wavelengths
excitations (they correspond to sound waves) and short wavelengths
(associated to free particle behavior) \cite{Stringaribook1}. In
other words, concerning the scale of length of these excitations the
Bogoliubov method has no restrictions at all. In connection with the
hydrodynamic analysis there are several approximations which are
introduced, and clearly stated in the literature. For instance, the
kinetic term appearing in the Gross--Pitaevski equation is separated
into two contributions \cite{Pethickbook}, namely, one related to
the motion of particle and, a second one, corresponding to
zero--point motion. The latter, known as quantum pressure term is
neglected, as a consequence of the Thomas--Fermi approximation. and,
therefore, the hydrodynamic analogue of the Gross--Pitaevski
equation is obtained. This approach, relying on the Thomas--Fermi
approximation, imposes a stringent condition upon the order of
magnitude of the possible wavelengths of the excitations. Indeed,
this model entails the fact that the kinetic energy is negligible
compared against the other energies of the problem, a point that
implies that all phenomena related to scale lengths smaller than the
so--called healing length \cite{Pethickbook} lie outside the
validity realm of the hydrodynamic approach. In other words, the
hydrodynamic model cannot provide us information about the case of
wavelengths smaller than the healing length, i.e., only large
wavelengths can be studied.

In addition, the deduction of the hydrodynamic equations assumes an
additional restriction, the one does not appear, usually, in the
literature, and that we, here, for the sake of completeness analyze.
In order to explain this point in the clearest way we will resort to
a usual deduction of this analogy, see  pages 167 to 169 of
\cite{Pethickbook}. We start with the time--dependent
Gross--Pitaevski equation

\begin{eqnarray}
i\hbar\frac{\partial\psi(\vec{r}, t)}{\partial t} =
-\frac{\hbar^2}{2m}\nabla^2\psi(\vec{r}, t)+
V_t(\vec{r})\psi(\vec{r}, t) +\nonumber\\
U_0\vert\psi(\vec{r}, t)\vert^2\psi(\vec{r},t) \label{equa1}.
\end{eqnarray}

In this last expression $V_t(\vec{r})$ denotes the trap used for the
confinement of the system, which for this particular case is given
by

\begin{eqnarray}
V_t(\vec{r})= \frac{m\omega_0r^2}{2} \label{equa2}.
\end{eqnarray}

Afterwards the following transformation is introduced in the
Gross--Pitaevski equation

\begin{eqnarray}
\psi(\vec{r},t)=
\sqrt{n(\vec{r},t)}\exp{(i\phi(\vec{r},t))}\label{equa3}.
\end{eqnarray}

In this last expression $n(\vec{r},t)$ and $\phi(\vec{r},t)$ are
real--valued functions. Then, we obtain two equations, one for the
real part of Gross--Piatevski and the second one for the imaginary
term.

\begin{eqnarray}
\frac{\partial n^2}{\partial
t}=-\frac{\hbar}{m}\nabla\cdot\Bigl(n^2\nabla\phi\Bigr)
\label{equa4},
\end{eqnarray}

\begin{eqnarray}
-\hbar\frac{\partial\phi}{\partial
t}=-\frac{\hbar^2}{2m\sqrt{n}}\nabla^2\sqrt{n}+\frac{1}{2}mv^2+ V_t+
nU_0 \label{equa5},
\end{eqnarray}

\begin{eqnarray}
\vec{v}= \frac{\hbar}{m}\nabla\phi, ~~~v=\vert\vec{v}\vert
\label{equa6}.
\end{eqnarray}

We now take the gradient of the last expression and obtain the
motion equation for the velocity, namely,

\begin{eqnarray}
m\frac{\partial\vec{v}}{\partial t}=-\nabla\Bigl(V_t+
nU_0+\frac{1}{2}mv^2-
\frac{\hbar^2}{2m\sqrt{n}}\nabla^2(\sqrt{n})\Bigr) \label{equa7}.
\end{eqnarray}

At this point the consequences of the Thomas--Fermi approximation
are introduced, in the sense that the kinetic energy is neglected
and therefore the quantum pressure term
($\frac{\hbar^2}{2m\sqrt{n}}\nabla^2\sqrt{n}+$) is discarded. We now
cast this last equation in a different way, the idea is to introduce
the concept of pressure in our formalism. In the case of a uniform
Bose gas the density $n$ is a constant and, therefore, the energy of
the system is $E=(N-1)NU_0/(2V)$, here $N>>>1$ is the number of
particles, whereas $V$ is the corresponding volume. Since
$p=-\frac{\partial E}{\partial V}$ then for the homogeneous case
$p=\frac{n^2U_0}{2}$, a fact that implies a motion equation for the
velocity with the following structure

\begin{eqnarray}
m\frac{\partial\vec{v}}{\partial
t}=-\nabla\Bigl(V_t+\frac{1}{2}mv^2)-
2\nabla\Bigl(\frac{p}{n}\Bigr)\label{equa8}.
\end{eqnarray}

Since $n$ is constant then $\nabla\Bigl(\frac{p}{n}\Bigr)=
\frac{1}{n}\nabla p$. But this expression is generalized to those
cases in which the density ($n$) is not a constant, a fact that
entails an approximation. In addition, the usual analogy (see,
equation (7.24) in \cite{Pethickbook}) assumes that the energy of
the condensate is the same as that emerging in a homogeneous gas.
Indeed, if we, within the Thomas--Fermi approximation, consider the
energy of the condensate, then we must include not only the mean
field energy ($E=N^2U_0/(2V)$) but also the energy stemming from the
trapping potential. We may estimate this last energy as follows: the
energy of one of the particles of the gas, due to the interaction
with the isotropic harmonic oscillator, is $m\omega^2_0R^2/2$, here
$R$ denotes the size of the condensate. Clearly, $V=R^3$, therefore,
for $N$ particles the whole energy reads
$E=N^2U_0/(2V)+Nm\omega^2_0V^{2/3}/2$. Then the pressure becomes
$p=n^2U_0/2 -nm\omega^2V^{2/3}/3$. Therefore, the correct
expression, containing the pressure, is given by

\begin{eqnarray}
m\frac{\partial\vec{v}}{\partial
t}=-\nabla\Bigl(V_t+\frac{1}{2}mv^2\Bigr)- \nabla\Bigl(2\frac{p}{n}
+ \frac{2m\omega^2V^{2/3}}{3}\Bigr) \label{equa9}.
\end{eqnarray}

This last expression becomes

\begin{eqnarray}
m\frac{\partial\vec{v}}{\partial
t}=-\nabla\Bigl(V_t+\frac{1}{2}mv^2\Bigr)- \frac{2}{n}\nabla p
+\frac{2p}{n^2}\nabla n \label{equa10}.
\end{eqnarray}

The approximation introduced in the case of non--homogeneous gases
is then

\begin{eqnarray}
\frac{2p}{n^2}\nabla n = \frac{1}{n}\nabla p \label{equa11}.
\end{eqnarray}

This last condition defines a functional dependence for $p$ as a
function of $n$.

\begin{eqnarray}
p\sim n^2.
\end{eqnarray}

It is a rough approximation, remember that previously it was found
that $p=n^2U_0/2 -nm\omega^2V^{2/3}/3$. In terms of characteristic
lengths we may state that this approximation implies that the
distance over which the pressure has a meaningful change is twice
the corresponding distance for the density. This last statement can
be understood noting that the approximation implies $n^2U_0/2
>>>nm\omega^2V^{2/3}/3$. We now recall that under the presence of mean
field interaction \cite{Uedabook} the size of the condensate is
given by $R=(Na/\tilde{l})^{1/5}\tilde{l}$, where $a$ is the
scattering length and $\tilde{l}=\sqrt{\hbar/(m\omega_0)}$. Joining
these last two conditions we find that it is a good approximation at
those points where the density
$n>>(Na/\tilde{l})^{2/5}/(a\tilde{l}^2)$. Therefore, at those points
where the density becomes smaller than
$(Na/\tilde{l})^{2/5}/(a\tilde{l}^2)$ the approximation is not a
good assumption. We now estimate this fact resorting to the
Thomas--Fermi condition in which $n=m\omega_0(R^2-r^2)/U_0$
\cite{Pethickbook}. Under these conditions we find that the validity
of the approximation implies that $\tilde{l}^2(1-8\pi)>r^2$, which
is not possible. The conclusion is that it is not a good assumption
for the case of an isotropic harmonic oscillator. These arguments
show that the analogy between the condensate at $T=0$ and
hydrodynamics contains not only the Thomas--Fermi approximation but
additional assumptions.

A careful look at the Bogoliubov equations \cite{Stringaribook1}
allows us to state that there are no assumptions as those implicit
in the hydrodynamic formalism. In other words, the solutions
obtained from the former model will provide a better description
than those stemming from the latter.

These arguments provide a motivation for the quest of solutions
resorting to Bogoliubov equations. This is the issue addressed here.
The corresponding solutions will be found and it will be shown that
for those cases in which the angular momentum does not vanish there
are an infinite number of frequencies for the elementary
excitations, just as in the hydrodynamic model \cite{Pethickbook}.
In addition, in contrast to the known situation \cite{Stringari1},
it will be shown that the mathematical structure of the our
solutions is not a polynomial, but an infinite series. Of course,
the convergence neighborhood is analyzed as a function of the
properties of the condensate.

One of our main results will be related to the way in which the
single--particle properties impinge upon the features of the
elementary excitations. Indeed, each one of the particles conforming
the gas (the depletion term is here neglected) lies, due to the
lowness of the temperature, in the ground state of its one--particle
Hamiltonian, i.e., a state with vanishing angular momentum. When the
interaction due to the mean field contribution (this term measures
the strength of the interaction that a particle in the BEC
experiences as a consequence of the presence of the remaining
particles in the gas) is not strong enough then the characteristics
of the frequencies of the elementary excitations can acquire only
those features belonging to the single--particle realm. The
threshold defining {\it not enough} is also analyzed.

\section{\bf Bogoliubov Equations}

The starting point is the time--dependent Gross--Pitaevski equation

\begin{eqnarray}
i\hbar\frac{\partial{\psi(\vec{r}, t)}}{\partial{t}} =
-\frac{\hbar^2}{2m}\nabla^2\psi(\vec{r},t)+
m\omega_0^2r^2\psi(\vec{r},t)/2 \nonumber\\+
U_0\vert\psi(\vec{r},t)\vert^2\psi(\vec{r},t) \label{equa12}.
\end{eqnarray}

Now a change in the order parameter ($\psi(\vec{r}, t)$) is
introduced in this last equation, namely,

\begin{eqnarray}
\delta\psi(\vec{r},t))=
\exp{(-it\mu/\hbar)}[u(\vec{r})\exp{(-i\omega t)}\nonumber\\
-{v}^{\star}(\vec{r})\exp{(+i\omega t})] \label{equa13}.
\end{eqnarray}

Then we obtain two equations, known as Bogoliubov equations
\cite{Bogoliubov1, Stringaribook1}.

\begin{eqnarray}
-\frac{\hbar^2}{2m}\nabla^2+ 2nU_0
-\hbar\omega-\mu+m\omega^2_0r^2/2]u(\vec{r})\nonumber\\
=[\mu-m\omega^2_0r^2/2]v(\vec{r}) \label{equa14},
\end{eqnarray}

\begin{eqnarray}
-\frac{\hbar^2}{2m}\nabla^2+ 2nU_0
+\hbar\omega-\mu+m\omega^2_0r^2/2]v(\vec{r})\nonumber\\
=[\mu-m\omega^2_0r^2/2]u(\vec{r}) \label{equa15}.
\end{eqnarray}

They are two coupled differential equations, and here we proceed to
uncouple them. Defining

\begin{eqnarray}
\hat{L}=-\frac{\hbar^2}{m}\nabla^2,\nonumber\\
\hat{H}=\mu-m\omega^2_0r^2,\nonumber\\
f(\vec{r})=v(\vec{r})+u(\vec{r}),\nonumber\\
F(\vec{r})=u(\vec{r})-v(\vec{r}) \label{equa16}.
\end{eqnarray}

We may cast the Bogoliubov equations as

\begin{eqnarray}\hat{L}f(\vec{r})=\hbar\omega F(\vec{r}) \label{equa17},
\end{eqnarray}

\begin{eqnarray}
\hat{L}F(\vec{r})=\hbar\omega f(\vec{r})-2\hat{H}F(\vec{r})
\label{equa18}.
\end{eqnarray}

Finally, from these last two equations we find for our function
$F(\vec{r})$ the following condition

\begin{eqnarray}
\hat{L}[\hat{L}F(\vec{r})+2\hat{H}F(\vec{r})]
=(\hbar\omega)^2F(\vec{r}) \label{equa19}.
\end{eqnarray}

The structure of our two equations allow us to seek our solutions
with the form

\begin{eqnarray}
F(\vec{r})_{(l)}^{(\tilde{m})}=
R(r)_{(l)}Y(\theta,\phi)_{(l)}^{(\tilde{m})} \label{equa20}.
\end{eqnarray}

Here$Y(\theta,\phi)_{(l)}^{(\tilde{m})}$ denote the spherical
harmonics and, in consequence, the parameters $l$ and $\tilde{m}$
are the angular momentum of the excitation and its $z$--component,
respectively. This last assumption allows us to end up with a
fourth--order differential equation.

\begin{eqnarray}
\frac{\hbar^4}{4m^2}[\frac{d^4R_{(l)}}{dr^4}+\frac{4}{r}\frac{d^3R_{(l)}}{dr^3}]\nonumber\\
+\frac{\hbar^2}{2m}[m\omega^2_0r^2-2\mu-\frac{l(l+1)\hbar^2}{2mr^2}]\frac{d^2R_{(l)}}{dr^2}\nonumber\\+
\frac{\hbar^2}{2m}[6m\omega^2_0r-\frac{4\mu}{r}+\frac{l(l+1)\hbar^2}{mr^3}]\frac{dR_{(l)}}{dr}\nonumber\\
+\frac{\hbar^2}{2m}[6m\omega^2_0-\frac{l(l+1)\hbar^2}{2mr^4}]R_{(l)}=(\hbar\omega)^2R_{(l)}
\label{equa21}.
\end{eqnarray}

At this point already a remarkable difference emerges when we
compare this last equation against the fundamental expression within
the hydrodynamic formalism. Indeed, the motion equation for the
elementary excitations (as a function of the perturbations in the
density, i.e., $\delta n$) is a second--order differential equation
\cite{Pethickbook}

\begin{eqnarray}
m\frac{\partial^2\delta n}{\partial t^2}=
\frac{1}{m}[\nabla(m\omega^2_0r^2/2)\cdot\nabla\delta n\nonumber\\
-(\mu-m\omega^2_0r^2/2)\nabla^2\delta n \label{equa22}.
\end{eqnarray}

\section{\bf Solutions to Bogoliubov Equations}

Our main result is contained in (\ref{equa21}) and now we take the
simplest situation, namely, $l=0$. It is readily seen that a
solution exist if

\begin{eqnarray}
R_{(l=0)}= constant,~~\omega= \sqrt{3}\omega_0 \label{equa23}.
\end{eqnarray}

The corresponding structure for the elementary excitations is

\begin{eqnarray}
\delta\psi(\vec{r},t))=\nonumber\\
\exp{(-it\mu/\hbar)}\frac{R_{(l=0)}}{2}\Bigl[\bigl(1+g(r)\bigr)\exp{(-i\sqrt{3}t\omega_0)}-\nonumber\\
\bigl(-1+g(r)\bigr)\exp{(i\sqrt{3}t\omega_0)}\Bigr] \label{equa24},
\end{eqnarray}

\begin{eqnarray}
g(r)=\frac{2\mu}{\sqrt{3}\hbar\omega_0}-
\frac{1}{\sqrt{3}}\Bigl(\frac{r}{\tilde{l}}\Bigr)^2 \label{equa25},
\end{eqnarray}

\begin{eqnarray}
\tilde{l}=\sqrt{\frac{\hbar}{m\omega_0}} \label{equa251}.
\end{eqnarray}

We now proceed to consider the solution for $l>0$. This will be done
in a different spirit to the analysis of the hydrodynamic situation.
In the latter the allowed frequencies are deduced imposing a
stringent condition upon the general solution of the corresponding
differential equation. In the case of the hydrodynamic model the
corresponding equation, for $l\not=0$, is the hypergeometric
equation \cite{Pethickbook}, and it is reduced to a polynomial
\cite{Stringari1} imposing some conditions upon the parameters of
the solution, see page 181 \cite{Pethickbook}. Of course, the
reduction of the hypergeometric function to a polynomial guarantees
the convergence of the solution at all points inside the condensate.
This procedure renders the frequencies for the excitations as a
function of the angular momentum (values of $l$) and the number of
radial nodes of the emerging polynomials, see expression (7.71) page
181 \cite{Pethickbook}. It must be stressed that the reduction of
the hypergeometric function to a polynomial is a sufficiency
condition for the convergence of the solution, though not a
necessity condition.

Here we consider the solution as an infinite series and look for the
conditions that ensure the convergence of the solution at all points
inside the condensate.

Our solution has the form (here the angular momentum does not
vanish, i.e., $l=1,2,...$)

\begin{eqnarray}
R_{(l)}= \sum_{s}b_sr^s \label{equa26}.
\end{eqnarray}

Introducing this function into (\ref{equa21}) we obtain the
following conditions

\begin{eqnarray}
l(l+1) b_0= 0\label{equa27},
\end{eqnarray}

\begin{eqnarray}
l(l+1) b_2= 0\label{equa28},
\end{eqnarray}

\begin{eqnarray}
\frac{\hbar^2}{m}\Bigl(6-l(l+1)\Bigr)b_3=2\mu b_1 \label{equa29},
\end{eqnarray}

\begin{eqnarray}
\frac{\hbar^4}{4m^2}\Bigl([s+5][s+4][s+3][s+2]-\nonumber\\
l(l+1)[s+3]^2\Bigr)b_{(s+4)}-\frac{\hbar^2\mu}{m}[s+3][s+2]b_{(s+2)}\nonumber\\
+\hbar^2\Bigl(\omega^2_0[s+3][s+2]-\omega\Bigr)b_{(s)},
~~s=0,1,2,...\label{equa30},
\end{eqnarray}

Concerning (\ref{equa27})--(\ref{equa29}) we  must add that they are
imposed to discard all kind of singularities. The last one is the
recursion relation for the coefficients of our solution. Notice that
these last conditions imply that all the coefficients of the type
$b_{(2s)}$ must vanish. In other words, in contrast with the
situation of the hydrodynamic formalism, where the solutions are
even functions of the radial coordinate (\cite{Stringari1}), here
only odd powers of $r$ emerge.

In order to illustrate the behavior of our solutions we analyze the
case $l=2$. Under this condition (\ref{equa27})--(\ref{equa30})
imply

\begin{eqnarray}
b_1= 0\label{equa31},
\end{eqnarray}

\begin{eqnarray}
b_5= \frac{48m\mu}{264\hbar^2}b_3\label{equa32}.
\end{eqnarray}

From the recursion condition (\ref{equa30}) we find the allowed
frequencies. Indeed, notice that this expression allows us consider
an $\omega$ for each $s=3,5,7,....$ as follows

Consider (\ref{equa30}) for $s=3$

\begin{eqnarray} \frac{366\hbar^4}{4m^2}b_{(7)} -\frac{30\hbar^2\mu}{m}b_{(5)}+\hbar^2\Bigl(30\omega^2_0-\omega\Bigr)b_{(3)}=0.\label{equa33}
\end{eqnarray}

Here we impose the condition

\begin{eqnarray}
30\omega^2_0-\omega\label{equa34}.
\end{eqnarray}

Then (\ref{equa33}) becomes

\begin{eqnarray}
b_{(7)}=\frac{30\mu}{366\hbar^2m}b_{(5)}.\label{equa35}
\end{eqnarray}

In other words, we obtain the frequency and the solution, with just
one free parameter, $b_{(3)}$

\begin{eqnarray}
R_{(l=2)}= b_{(3)}r^3\Bigl[1+
\frac{48}{364}\frac{m\mu}{\hbar^2}r^2\nonumber\\
+\frac{48}{364}\frac{30}{264}(\frac{m\mu}{\hbar^2})^2r^4+...
\Bigr].\label{equa36}
\end{eqnarray}

Let us now address the issue concerning the convergence of this
series. Our expressions imply that

\begin{eqnarray}
b_{(2s+1)}=(10^{-1})^{s}(\frac{m\mu}{\hbar^2})^{s-3}\Bigl[(\frac{m\mu}{\hbar^2})^2\nonumber\\
-\frac{1}{10\tilde{l}^4}\Bigr]b_{(3)}.\label{equa37}
\end{eqnarray}

A necessary but not sufficient condition \cite{Takeuchibook1} for
the convergence of this series reads

\begin{eqnarray}
b_{(2s+3)}r^{(2s+3)}/(b_{(2s+1)}r^{(2s+1)})\rightarrow 0, ~~if~~
s\rightarrow\infty.\label{equa38}
\end{eqnarray}

For our case this condition entails

\begin{eqnarray}
r^2<\frac{10\hbar^2}{m\mu}.\label{equa38}
\end{eqnarray}

In order to evaluate this condition in terms of the parameters of
the condensate let us consider the value of the chemical potential
according to the Thomas--Fermi approximation ($\mu =
m\omega^2_0R^2/2$ with the relation
$R=(Na/\tilde{l})^{1/5}\tilde{l}$ \cite{Pethickbook}). Under these
conditions (\ref{equa38}) becomes

\begin{eqnarray}
r<\sqrt{20}\Bigl(\frac{\tilde{l}}{Na}\Bigr)^{1/5}\tilde{l}.\label{equa39}
\end{eqnarray}

A convergence radius, at least equal to $R$, requires

\begin{eqnarray}
R\leq r.\label{equa39}
\end{eqnarray}

Therefore, our solution is valid in the whole condensate if

\begin{eqnarray}
(Na/\tilde{l})^{2/5}<\sqrt{20}.\label{equa41}
\end{eqnarray}

This last expression provides the necessity requirements needed to
have a convergent series at all points within the condensate, as a
function of its parameters.

The sufficiency condition that guarantees the convergence of our
solution requires \cite{Takeuchibook1} that $\forall\epsilon>0$,
there exists an integer $M$ such that

\begin{eqnarray}
\sum_{s=M}^{\infty}b_{(2s+1)}r^{(2s+1)}<\epsilon.\label{equa42}
\end{eqnarray}

We may obtain an upper bound for this expression, since
(\ref{equa38}) is fulfilled.

\begin{eqnarray}
\sum_{s=M}^{\infty}b_{(2s+1)}r^{(2s+1)}\leq\frac{\Bigl(\frac{m\mu
r^2}{10\hbar^2}\Bigr)^{M-1}}{1-\frac{m\mu
r^2}{10\hbar^2}}b_{(3)}.\label{equa43}
\end{eqnarray}

It readily seen that we may find an integer $M$ such that

\begin{eqnarray}
\Bigl(\frac{m\mu
r^2}{10\hbar^2}\Bigr)^{M-1}<\frac{\epsilon}{b_{(3)}}\Bigl(1-\frac{m\mu
r^2}{10\hbar^2}\Bigr) .\label{equa44}
\end{eqnarray}

Inserting (\ref{equa44}) into (\ref{equa43}) we conclude that the
series does converge. In other words, for this particular situation,
the necessity condition turns out to be a sufficiency condition.

Additional frequencies and solutions can be found as follows. The
frequency has been obtained, in the present case, from the recursion
expression (\ref{equa30}) imposing the condition
$\omega^2_0(s+3)(s+2)-\omega^2=0$, for $s=3$. More frequencies
emerge if we demand the fulfillment of this condition for $s=4, 5,6
...$. The convergence of the corresponding series is guaranteed by
the structure of the recursion expression.

\section{\bf Bogoliubov Equations versus Hydrodynamic Formalism}

Notice that our solution for $l=0$ (see expressions (\ref{equa24})
and (\ref{equa25})) describes a mode which is more localized near
the surface of the cloud, i.e., it corresponds to surface waves. The
mathematical structure of our solutions, series in the radial
coordinate, tell us that this last characteristic of the case $l=0$
is shared by all the solutions, namely, they are related to surface
waves. We may then state that for particle--like modes the
elementary excitations of a BEC trapped by an isotropic harmonic
oscillator are surface waves. Of course, due to the spherical
symmetry of the solutions we are in the presence of degenerated
waves. In addition, the higher the frequency becomes, the more
localized near the surface that these waves are.

Our formalism for the case of vanishing angular momentum does not
coincide with the results of the hydrodynamic procedure, in which
the frequency is given by $\omega(s,l)=\omega_0(2s^2 +2sl
+3s+l)^{1/2}$, here $s$ denotes the number of radial nodes of the
solution \cite{Stringari1}, i.e., it has an infinite number of
frequencies related to the case $l=0$. If $l=0$, then
$\omega(s,0)=\omega_0(2s^2+3s)^{1/2}$.

Our frequency, just one possible value, is given by
$\omega=\sqrt{3}\omega_0$, a result which does not coincide with any
of the possible cases of the hydrodynamic formalism. As a matter of
fact, our frequency is smaller than all the possible values of the
other model, whose smallest value reads ($s=0$ is discarded in this
discussion) $\omega(s=1,l=0)=\sqrt{5}\omega_0$. In other words, for
vanishing angular momentum our assertion is that collective modes
have always larger frequencies than those of other possible modes.

We may explain this fact as a consequence of the role that the
kinetic energy term plays in the definition of the dynamics of these
elementary excitations. Indeed, resorting to a sum rule approach
\cite{Lipparini1} we find that the introduction in the hydrodynamic
procedure of the effects of the kinetic energy term produces a
reduction of the frequency the one depends upon the ratio of the
energy related to the harmonic oscillator ($E_{(ho)}$) and the
kinetic energy ($E_{(kin)}$) \cite{Stringari1}, which for the case
of $s=1$ and vanishing angular momentum takes the form
$\omega^2=\omega^2_0(5-E_{(kin)}/E_{(ho)})$. This last fact explains
our result, indeed, since our method does not discard the effects of
the kinetic energy we should expect a lower frequency than that
related to the smallest frequency of the hydrodynamic formalism. In
this case it corresponds to $E_{(kin)}/E_{(ho)})= 2$, a fact that
confirms that we are outside the validity region of the
Thomas--Fermi approximation.

Another interesting point of our result stems from the fact that the
frequencies are $l$--independent, see (\ref{equa30}), though the
solutions do depend upon the value of the angular momentum. This
dependence can be seen at the recursion relation where $l$ appears
explicitly. In the solutions of the hydrodynamic model both,
frequencies and solutions, show an explicitly dependence upon the
corresponding value of the angular momentum.

We may interpret this results as follows. The smallness of the
parameter $(Na/\tilde{l})^{1/5}$, required for the convergence of
our solutions ($(Na/\tilde{l})^{2/5}<\sqrt{20}$), entails that the
strength of the repulsive interaction (codified here in the factor
$Na$) is not strong enough in order to allow the frequency of the
elementary excitations to be determined as a function of the angular
momentum of the proposed solution (\ref{equa20}). In other words,
the presence of the trap and of the scattering length imply that the
ensuing solutions depend upon these aforementioned parameters,
though the frequency does depend upon the angular momentum (as
happens in the hydrodynamic approach) only if the parameter
$(Na/\tilde{l})^{1/5}$ is beyond a certain threshold, which for our
case can be estimated to be $\sqrt{20}\leq(Na/\tilde{l})^{2/5}$. In
other words, it is the repulsive interaction the one responsible for
the appearance in the frequency of the elementary excitations of the
angular momentum of the corresponding solution. We may then
generalize this last conclusion as a conjecture: In a trapped BEC
the frequency of the elementary excitations acquire information of
the angular momentum of the solution only by means of the parameter
$(Na/\tilde{l})^{1/5}$, and this happens only if this parameter is
beyond a certain threshold, in the present case it is,
approximately, $\sqrt{20}\leq(Na/\tilde{l})^{2/5}$. This last
condition can be understood from a different perspective. Indeed,
each one of the particles conforming the gas (the depletion term is
here neglected) lies, due to the lowness of the temperature, in the
ground state of its one--particle Hamiltonian, i.e., a state with
vanishing angular momentum. Since the interaction due to the mean
field contribution (this term measures the strength of the
interaction that a particle in the BEC experiences as a consequence
of the presence of the remaining particles in the gas) is not strong
enough then the characteristics of the frequencies of the elementary
excitations can depend only from those features belonging to the
single--particle realm. A further argument in this direction can be
seen in the fact that in the hydrodynamic approach the case $l=1$
corresponds to a translation of the cloud with no change in its
internal structure \cite{Pethickbook}. This kind of motion involves
a bulk movement of the particles of the BEC. Notice that in our case
this kind of effect is absent, and the reason lies in the fact that
the repulsive interaction is not strong enough and, therefore, no
bulk properties can emerge.

If we consider any value of the allowed frequency associated to the
case here explicitly shown ($l=2$) we find from the recursion
expression (\ref{equa30}) that it emerges for any other
non--vanishing value of $l$. Our  last comment also explains this
degeneracy of the frequencies, namely, they are obtained only from
the term whose coefficient is $\hbar^2$ in (\ref{equa30}), i.e., it
does not contain $l$.


\begin{thebibliography}{}
\bibitem{Stringari1} S. Stringari, Phys. Rev. Lett.\textbf{77},
2360 (1996).

\bibitem{Raman1} C. Raman, et al, Phys. Rev. Lett.\textbf{83},
2502 (1999).

\bibitem{Pethickbook}
 C. J.~Pethick and H.~Smith, \emph{Bose--Einstein Condensation in Dilute Gases} (Cambridge University Press,
 Cambridge, 2004).

\bibitem{Uedabook}
 M.~Ueda, \emph{Fundamentals and New Frontiers of Bose--Einstein Condensation} (World Scientic,
 Singapore, 2010).

\bibitem{Bogoliubov1} N. N. Bogoliubov, J. Phys. (Moscow) \textbf{11},
23 (1947).

\bibitem{Stringaribook1}
 L.~Pitaevsi and S.~Stringari, \emph{Bose--Einstein Condensation} (Oxford Science Publications,
 Oxford, 2003).

 \bibitem{Takeuchibook1}
Y.~Takeuchi, \emph{Sucesiones y Series} (Editorial Limusa,
 M\'exico, 1976).

\bibitem{Lipparini1} E. Piparrini and  S. Stringari, Phys. Rep. \textbf{175},
103 (1989).
\end{thebibliography}
\end{document}